\documentclass[preprintnumbers,amsmath,amssymb,twocolumn]{revtex4}
\usepackage{graphicx}% Include figure files
\usepackage{dcolumn}% Align table columns on decimal point
\usepackage{bm}% bold math

\begin{document}

\title{Electron spin resonance from NV centers \\ in diamonds levitating in an ion trap}

\author{T. Delord$^{1,2}$}
\author{L. Nicolas$^{1}$}
\author{L. Schwab$^{1}$}
\author{G. H\'etet$^{1}$} 

\affiliation{$^1$Laboratoire Pierre Aigrain, Ecole normale sup\'erieure, PSL Research University, CNRS, Universit\'e Pierre et Marie Curie, Sorbonne Universit\'es, Universit\'e Paris Diderot, Sorbonne Paris-Cit\'e, 24 rue Lhomond, 75231 Paris Cedex 05, France. \\
$^2$D\'epartement de Physique, \'Ecole Normale Sup\'erieure de Lyon, Universit\'e de Lyon, 46 All\'ee d'Italie, F-69364 Lyon, cedex 07,France
}

\begin{abstract}
We report observations of the Electron Spin Resonance (ESR) of Nitrogen Vacancy (NV) centers in diamonds that are levitating in an ion trap. Using a needle Paul trap operating under ambient conditions, we demonstrate efficient microwave driving of the electronic spin and show that the spin properties of deposited diamond particles measured by the ESR are retained in the Paul trap. We also exploit the ESR signal to show angle stability of single trapped mono-crystals, a necessary step towards spin-controlled levitating macroscopic objects. 
\end{abstract}
%
%\pacs{32.30.Bv, 37.10.Ty, 05.30.-d, 67.80.dj}
\maketitle

The negatively charged Nitrogen Vacancy (NV$^-$) center in diamond has emerged as a very efficient source of single photons and a promising 
candidate for quantum control and sensing via its electron spin. 
Recently, there has been much interest in the electronic spin of the NV$^-$ center in levitating diamonds \cite{Horowitz, Neukirch2}. This interest is partly motivated by proposals for hybrid optomechanics \cite{Rabl}, and implications in ultrahigh force sensitivity \cite{Kolkowitz} where 
the NV center's spin response to magnetic fields is exploited to read-out the motion of the diamond with high spatial resolution under ambient conditions \cite{Balasubramanian}. Amongst the many levitation schemes, optical traps are the most widely used \cite{Ashkin, Horowitz, Neukirch, Li}. They provide efficient localisation for neutral and charged particles and can work under liquid or atmospheric environnements. However the trap light that is scattered from the object means that excessive heating can be at work \cite{Ashkin, Hoang, Rahman, Neukirch}. Furthermore, optical traps may quench the fluorescence of NV centers \cite{Neukirch} and affect the electronic spin resonance contrast.

Being able to trap diamonds hosting NV centers without light scattering
could thus offer a better control of the spin-mechanical coupling and enlarge the range of applications of levitating diamonds. Levitation techniques such as ion traps \cite{Pau90} or magneto-gravitational traps \cite{Hsu} are tantalizing approaches for reaching this goal.
Ion traps could not only provide an escape route for scattering free trapping, but also enable a high localisation of the particles together with large trap depths as demonstrated by the impressive control over the motion that have been developped with single ions in the past \cite{Leibfried}. Various nano-objects have been confined in ion traps already, from coloidal nanocrystals \cite{Bell}, silica nanospheres \cite{Millen, Barker2010}, graphene flakes \cite{Nagornykh}, micron size diamond clusters containing NV centers \cite{Kuhlicke}, showing their potential for the motional control of macroscopic objects.

\begin{figure}[ht!]
\centerline{\scalebox{0.20}{\includegraphics{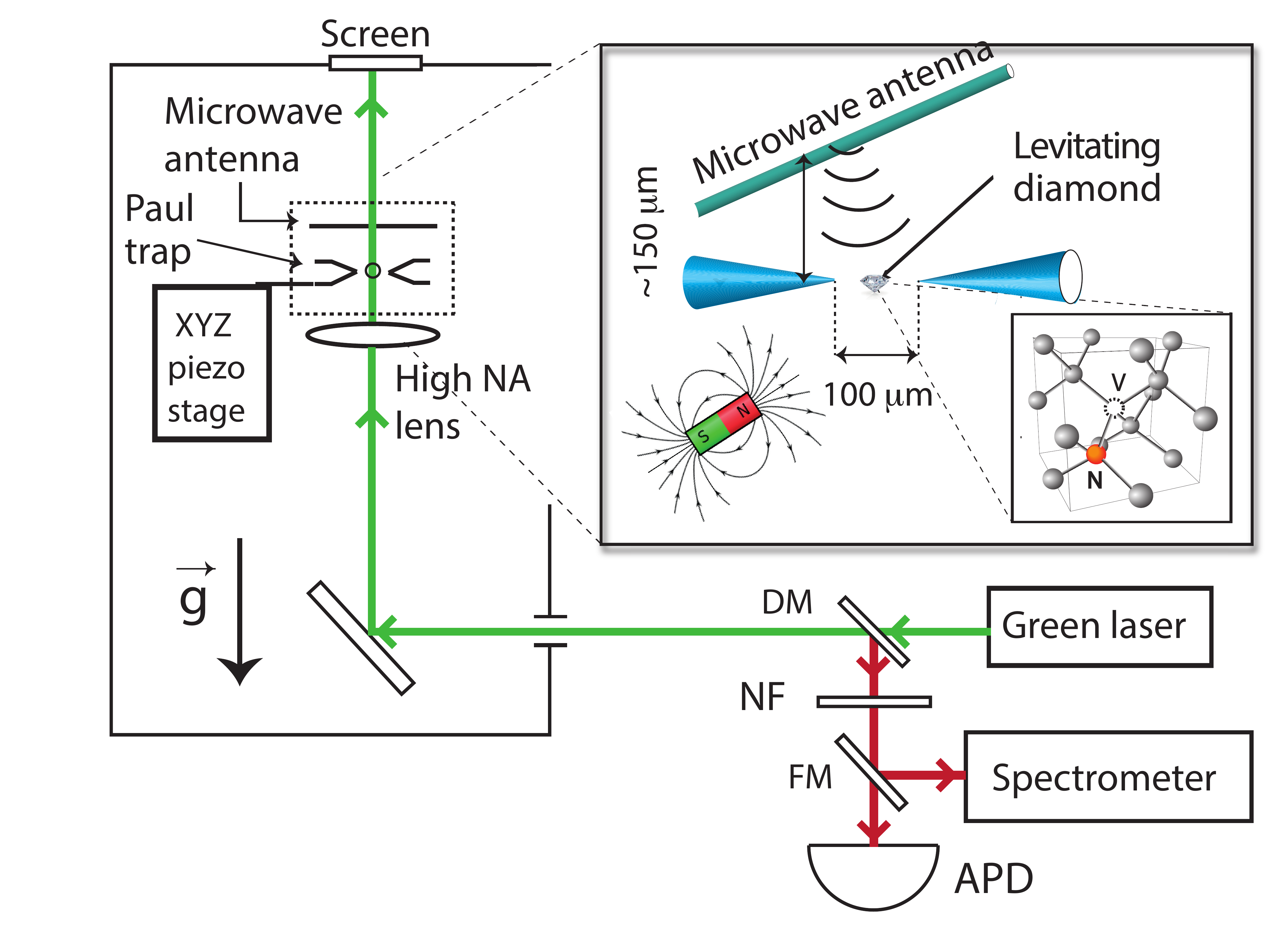}}}
\caption{Sketch of the optical setup. A green laser is focused onto a diamond levitating in the Paul trap.  The diamond position is monitored using phase contrast imaging on a distant screen and the photoluminescence from the NV centers in the trapped diamond is collected by the same objective and measured either on an avalanche photodiode (APD) or on a spectrometer. As shown in the inset, a microwave antenna is brought 150 $\mu$m away from the trap center to control the NV centers' electronic spins and a neodynium magnet is placed 5 cm away from the trap center to Zeeman shift the NV centers electronic spins. DM~: Dichroic mirror, FM~: flipping mirror, NF~: Notch filter centered at 532 nm.
}\label{Setup}
\end{figure}

In this work, we report measurements of the electronic spin resonance of NV centers embedded in diamonds that are levitating in an ion trap.
Further, we observe high contrast Zeeman-splitted levels, demonstrating angular stability over single levitating monocrystals on time scales of minutes, paving the way towards single spin opto-mechanical schemes in scattering-free traps.
 
 \subsection{The Paul trap}

An ion trap typically consists of electrodes that are placed at an oscillating potential generating a time-varying quadrupolar electric field. In the adiabatic regime, this provides a ponderomotive force that brings charged particles at a minimum of the electric field intensity \cite{Pau90}.
The trap that we use is a Paul-Straubel trap \cite{Straubel,Yu} operating under ambient conditions, and consisting of two tungsten needles with a radius of curvature of 25 $\mu$m,  
surrounded by an uncritical ground electrode structure.
The distance $d=2z_0$ between the needles is around $100~\mu$m. 
We operate the trap with a peak-to-peak voltage ranging from $V_{\rm ac}$=1000~V to 4000~V at $\Omega/2\pi\approx 5$ kHz.

The curvature of and distance between the needles crucially determine the confinement and the potential depth in both the radial and axial planes \cite{Deslauriers}. 
The axial angular frequency $\omega_z$ of the harmonic pseudo-potential is given by 
\begin{eqnarray}
\omega_z&=&\frac{|Q_{\rm tot}| V_{\rm ac} \eta}{\sqrt{2} m \Omega z_0^2},
\end{eqnarray}
where $V_{\rm ac}$ is the peak to peak voltage applied between the electrodes and the far distance surrounding ground, 
$m$ is the mass of the trapped particle and $\eta$ the efficiency factor
that accounts for the reduction in the trap potential as compared 
to an analogous quadrupole trap with hyperbolic electrodes.
% that have end caps spacing of $2 z_0$ and an inner diameter $2\sqrt{2}z_0$.
$\Omega/2\pi$ is the trapping frequency and $Q_{\rm tot}$ is the total excess charge of the particle. 
3D numerical simulations show that with our trap geometry and assuming charge to mass ratios on the order of mC.kg$^{-1}$ \cite{Kuhlicke} we can expect macro-motional frequencies $\omega_z$ that range from 100 Hz to several kHz, depending on $\Omega$ and $V_{\rm ac}$. 

The advantage of this needle trap over 3D linear traps is the very opened geometry. 
This means that high optical access is available for efficient collection of the photoluminescence of the embedded quantum emitters. In this work we use an aspherical lens with a numerical aperture of 0.77 at a working distance of 3.1 mm with no observable perturbations to the diamond motion due to the lens surface charges. Another advantage is flexibility :  since we can tune the distance from the trap center to the needles, we can reduce the capture volume after diamond loading, and thus increase the confinement {\it a posteriori} by bringing the two needles closer to each other. Similar trapping conditions in terms of confinement and optical access are feasible using ring, or planar traps but the needle trap geometry is the easiest to be tuned whilst particles are trapped.

Fig. \ref{Setup} shows a sketch of the experimental apparatus.
The position of the aspherical lens and the Paul trap is tuned using micropositioning stages.
One needle of the trap is also mounted on an XYZ piezo stage %(P-611.3 Nanocube from PI)
for fine tuning of the diamond to needle and diamond to laser distances. 
The lens, trap and micropositioning stages are then all enclosed in a box to minimize air currents and are controlled from outside.
The diamonds that we used are in the form of monocristalline powders (MSY micron-diamond powder from Microdiamant AG) and did not undergo specific processing. 
%\subsection{Microdiamond used : photoluminescence count rate and spectrum}
%Datasheet available at :\\
%http://www.microdiamant.com/products/micron-diamond-powder/monocrystalline-diamond-msy/ \\
They are produced by HPHT (high-pressure, high-temperature) synthesis and sold in the form of powders containing different mean particle sizes. We have observed NV centers in all the different diamond sizes $d$ we have studied, ranging from $d<$500 nm to $d\sim$12 $\mu m$. The NV density was found to be highly inhomogeneous from one diamond to another.
\begin{figure}[ht!]
\centerline{\scalebox{0.15}{\includegraphics{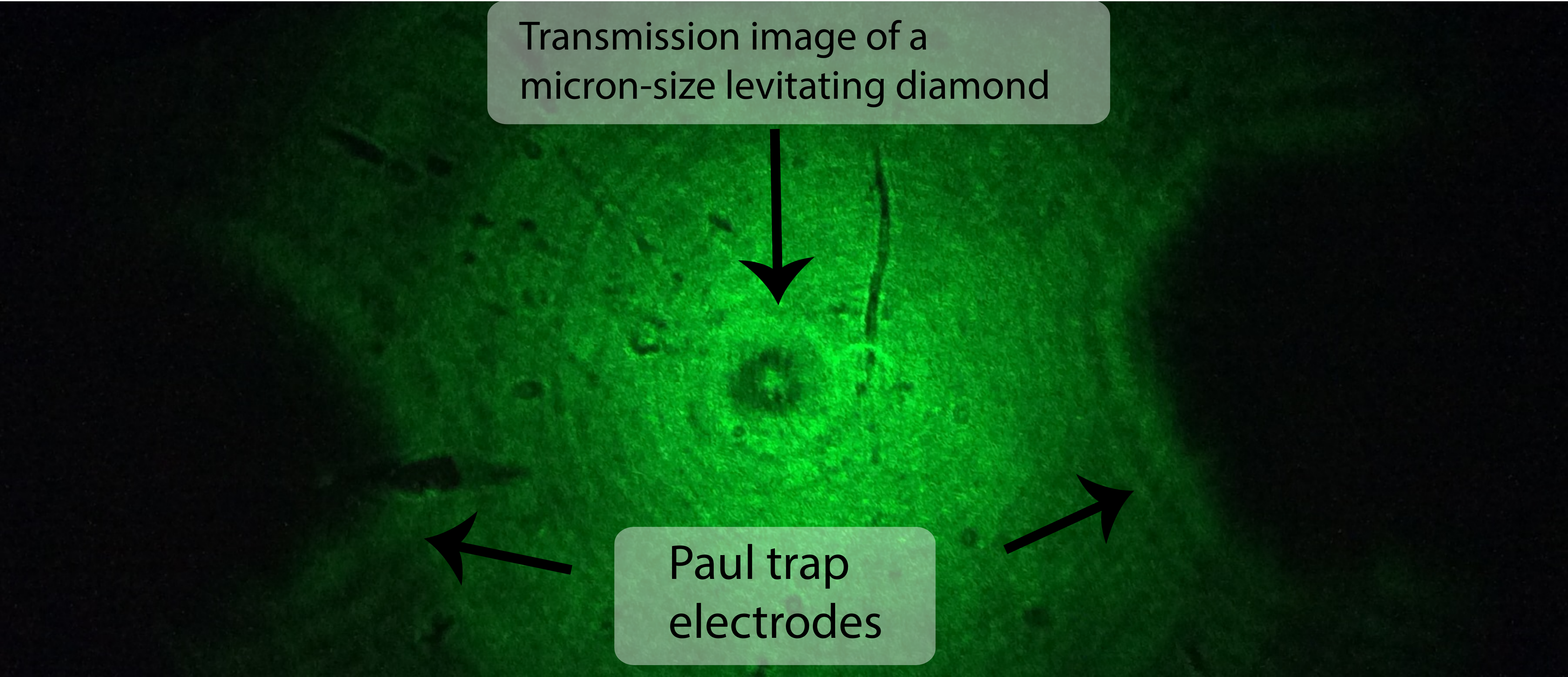}}}
\caption{Phase contrast imaging of a single levitating micro-diamond. The various spots on the lens are the diamonds that did not make it to the center of the trap and got deposited on the surface of the aspherical lens that lies 3.1 mm below the trap.
}\label{PCI}
\end{figure}

We load the diamond particles by first dipping a 300 $\mu$m copper wire in the powder and then bringing it in the vicinity of the trap center. 
Several mechanisms can be at work when the NDs are expelled from the copper wire. 
Electrons in the wire could be set in motion by the ponderomotive force of the Paul trap, reach the tip of wire and push the already charged diamonds towards the trap center. 
Another possibility is that charged diamonds are attracted to the trap center from the wire by the ponderomotive potential or by the static Coulomb force
itself. Further experiments would be needed to discriminate these effects. This loading method gives us a reliable and simple way to load the diamonds as each trapping run only requires one minute on average. 
This loading technique further provides a charge to mass ratio that enables trapping of the diamond particles for several days under ambient conditions. 
\begin{figure}[ht!]
\centerline{\scalebox{0.12}{\includegraphics{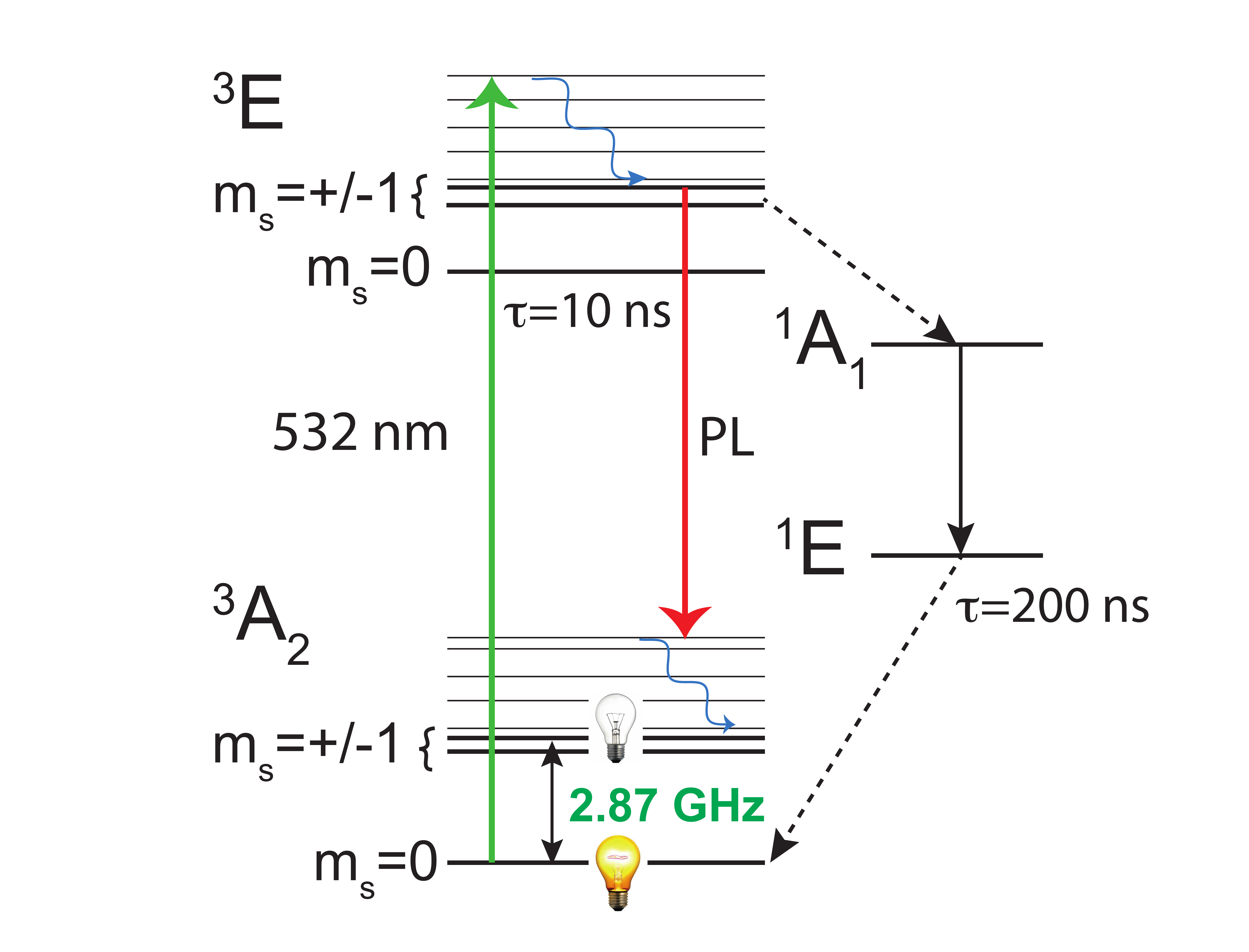}}}
\caption{The NV$^-$ center level scheme. Plain arrows describe radiative decay/absorptive channels and wavy lines represent phonon emission. Dashed lines are non-radiative decay paths.}
\label{level}
\end{figure}
The trapped particles are analysed using a green laser beam with a power ranging from 100~$\mu$W to 3.6 mW for phase contrast imaging (see Fig. 1) and for the photoluminescence measurements. To monitor the particle motion, we use the simplest form of phase contrast imaging : we measure the interference between the input laser field and the scattered field from the diamond. For our diamond sizes, the resulting image is the shadow of the diamond plus some interfering rings due to the sharp features on the side of the diamond, see figure \ref{PCI}.

Phase contrast imaging provides an efficient way to detect the micromotion and the eventual rotation of the particle on itself by measuring the diamond shadow position when the settings of the trap are changed. We could for instance deduce the trapping frequencies.
To measure them, we perform a slow ramp of the trapping frequency downwards from 4 kHz to a few kHz (the so called micromotional frequency). At resonance, when $\Omega$ is close to one of the macromotional frequencies $\omega$, the adiabatic condition (i.e. $\Omega\gg \omega$) is no longer fulfilled and the motion of the particle becomes unstable, thus providing a mean to measure $\omega$. This instability can be characterized using the $q$ parameter of the Mathieu equation \cite{Pau90} : the motion first becomes unstable for the most confining axis, at $q\geq q_{max}=0.908$.
Here the first observed resonance corresponds to the frequency $\omega_z$. It is typically measured to be about 1 kHz for diamonds that have a 10 $\mu$m diameter. Under usual working conditions ($\omega_z\ll \Omega$) this yields an effective macromotional frequency of several hundreds of Hz. Using the stability criterion and the diamond density, we can then extract the total charge on the surface \cite{Wuerker}.
The stability parameter $q_{max}$ relates to $|Q_{\rm tot}| / m$ via the formula 
\begin{eqnarray}
\frac{|Q_{\rm tot}|}{m}=\frac{q_{\rm max}}{ 4 \xi} \Omega^2,
\end{eqnarray}
where $\xi$ is the curvature of the static electric potential. 3D simulations of the electric potential of our trap gives us $\xi=2\times10^6~V/m^2$ and so the total number of elementary excess charges $|Q_{\rm tot}|$ on the surface of the diamond is about 5000.  
Note that this is only an estimate since damping due to collisions with gaz particles slightly modifies the instability and trap frequency \cite{Hasegawa1995}.   
We also measured the sign of the total charge by adding a constant voltage to the needles. Due to residual static electric fields, the particles are slightly off centered. Adding an extra DC voltage displaces particles away from or closer to the center of the trap depending on the relative sign between the voltage and surface charges. Applying a positive voltage on more than 20 different particles systematically displaced them towards the trap center, letting us conclude that the total surface charge is negative.

\subsection{Effect of the radiation pressure}

An important observation that was made is a pronounced rotation of the particles around the laser optical axis for diamonds that are below around 2 $\mu$m in diameter. This rotation takes place on time scales of milliseconds at power levels above hundreds of $\mu$W. For larger particles ($>2~\mu$m in diameter) and power levels below 300 $\mu$W, the particles appear to be stable on time scales of minutes.  Several measurements point to a radiation pressure induced perturbation of the diamond position due to the exciting laser. 

An order of magnitude of the laser induced force can be obtained by considering only a radiation pressure force along the optical axis for simplicity. To estimate the magnitude of the radiation pressure, we also consider the simplest case of a spherically shaped diamond particle at the focal point of a beam with a total power $P$ that uniformly fills the input lens. We denote $\theta_m$ the angle subtended by the lens. Here, we assume the particle size to be larger than the wavelength $\lambda$=532 nm, so that the ray optics approximation applies.

In our approximate case study, all rays are perpendicular to the surface of the particle so the mean momentum transfer per photon is $2 R_n \frac{h}{\lambda}$ where $R_n \sim 0.2$ is the Fresnel reflection coefficient for a normal wave and $h$ the Planck constant. Since the mean photon rate is $\frac{P \lambda}{h c}$, the total momentum transfer per unit time along the optical axis \textit{i.e.} the radiation pressure force is :
\begin{eqnarray}
\nonumber
F_{rad}&=&\int_{-\theta_m}^{\theta_m} \frac{h}{\lambda} 2 R_n \cos \theta \frac{P \lambda}{h c } \frac{d\theta}{ 2 \theta_m} \\
&=&  \frac{2 R_n P}{c} {\rm sinc} (\theta_m).
\end{eqnarray}

We compare this force to the force induced by the trap by considering the displacement $\Delta x$ of the particle from the center of the trap.  At equilibrium, we have 
$$\Delta x =\frac{ F_{rad}}{m \omega_x ^2}.$$

For a 10 $\mu$m diamond with a confinement frequency $\omega_x / 2 \pi  \sim 1$ kHz and a beam power $P=1$ mW, we get
$$\Delta x / P \sim 350~{\rm nm / mW}$$

For smaller diamonds ($\sim$ 2.8 $\mu$m), experiments were done with a slightly higher confinement frequency ($\omega_x / 2 \pi  \sim 1.5$ kHz) but the mass being smaller we obtain 
$$\Delta x / P \sim 11~\mu {\rm m / mW}$$

In the plane orthogonal to the optical axis $x$, since the trapped particles are not spherical, we actually expect the forces due to reflected photons not to cancel out. The resultant force is thus highly dependent on the shape and orientation of the particle.

Those numbers are in good agreement with visual observations of the apparent motion of the particle in the trap : for large particles, the particle is slightly displaced when the beam power is increased to a few mW while for smaller particles, a displacement from the trap center is observed with beam powers on the order of a few hundreds of 100 $\mu$W.
%Besides, a rotation of the particle about the optical axis and/or an important micromotion\\ 
Other qualitative observations point towards such a radiation pressure effect. We were for instance able to displace and/or to rotate deterministically single particles crystals : due to the Coulomb force, when many particles are in the trap, they are separated by more than microns and form a "Coulomb crystal". A laser can thus be applied to one side of this "multiparticle crystal" to make it rotate. 
A rotation was for instance observed when using three particles trapped in a plane perpendicular to the two needles when the laser beam- and thus the radiation pressure- was focussed on one side of the crystal.  Further studies will be conducted in order to better understand the motion of dynamically trapped particles in such moderately intense laser beams.

We now turn to the measurement of the photoluminescence from NV centers imbedded in the levitating diamond.

\subsection{Photoluminescence count rate and spectrum}

The photoluminescence (PL) of the NV centers is collected using the confocal microscope described in Figure~\ref{Setup}.
The green laser excites the NV centers in the phonon continuum (see level scheme in Fig. \ref{level}). 
The PL signal is then filtered using a Notch filter centered at 532 nm and can be directed either onto an avalanche photodiode or a spectrometer.
The NV$^-$ centers typically emit about 5\% of the total photoluminescence intensity in the zero-phonon line (ZPL) 
at 637 nm and 95\% of the PL into the phonon sidebands ranging from around 640 to 800 nm.
With a mW of laser excitation, we can collect around $10^5$ counts per second on the avalanche photodiode (SPCM-AQRH-15 from Perkin Elmer). 

A study of the photoluminescence of microdiamonds MSY 8-12 (9.6 $\mu m$ mean diameter) deposited on a glass plate has first been carried out in order to compare their properties with the one of trapped diamonds. Under up to 3.6 mW of 532 nm laser excitation with the same N.A.= 0.77 aspherical lenses for collection and excitation, the photoluminescence count rate detected by our multimode-fibered APD can be as high as 200 MHz for some diamonds. For other diamonds, the count rate can go bellow 0.5 MHz with the same excitation power. There is therefore a variation in the NV density by more than two orders of magnitude from one diamond to an other.

One should also point out that because we do not use aberration-corrected optics for the excitation and the collection of photoluminescence, there is a mismatch between the focal points of the excitation laser and the one corresponding to the optimum photoluminescence detection. For the presented experiment, we have chosen to optimize the input lens position in order to have an efficient collection, at the expense of excitation efficiency. Let us note that since many of the investigated diamonds are much larger than the excitation beam waist at the focal point, we could not saturate all the NV centers with this setup. We estimate the minimum number of NV centers that are excited to vary between a dozen to 4000 for 10 microns size diamonds. 

%A correction set up for the green laser comprising two lenses  -one on a translation stage- can and have occasionally been used to displace the green laser focus point at will so we can both excite and collect efficiently. Again, even if this improves the count rate, it will still not allow us to saturate all the NV centers of a micron-sized diamond. Given this facts, 

\begin{figure}
\includegraphics[scale=0.5]{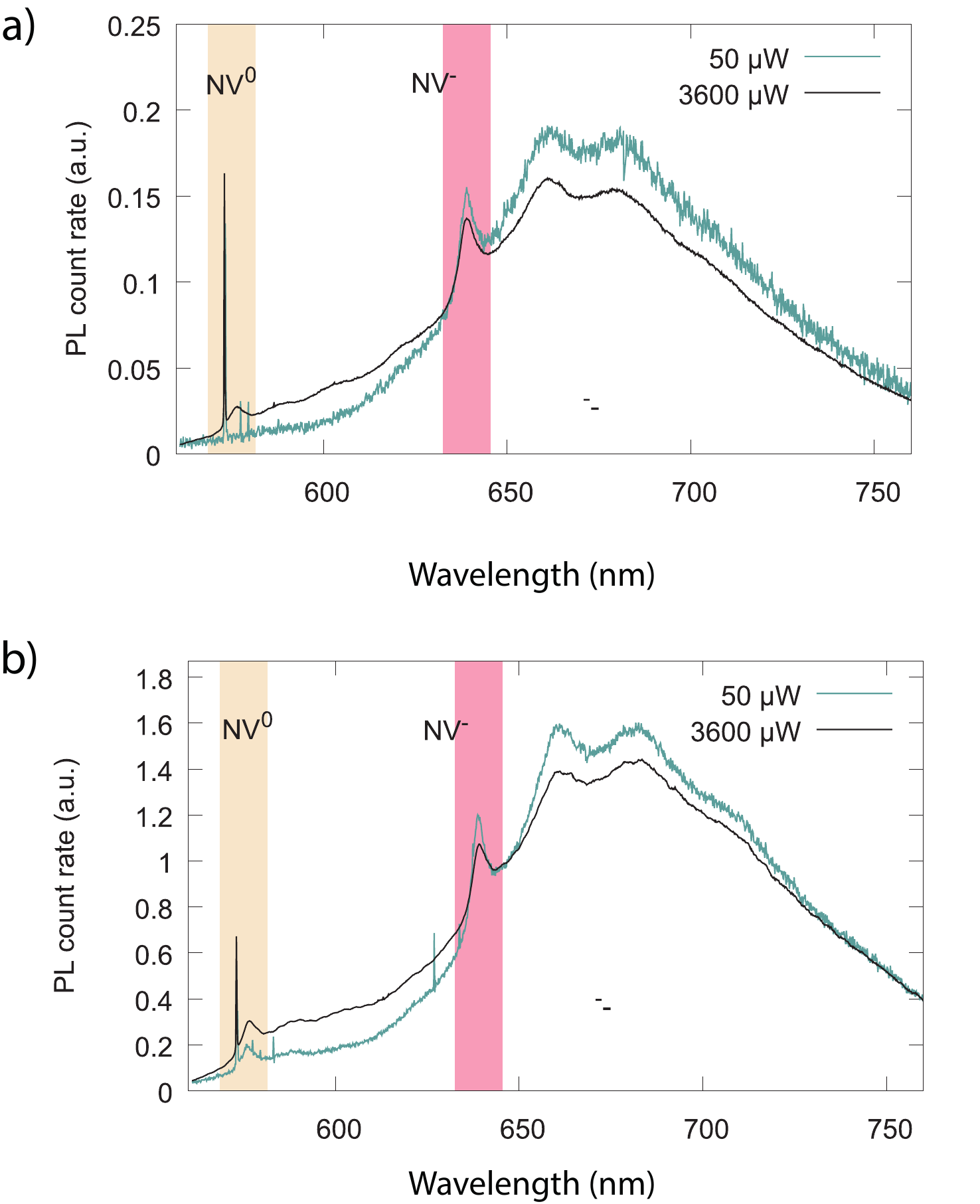}
%\\
%\end{tabular}
\caption{\label{spectrum}
Spectra obtained using two different excitation powers for 10 $\mu m$ sized diamonds deposited on a glass coverslip (trace (a)) and trapped in the Paul trap (trace (b)). The ZPL (zero-phonon line) of both the NV$^-$ and NV$^0$ centers are seen at 637 nm and 575 nm respectively.
The count rates have been corrected for
background light and normalized by the acquisition time and the excitation laser power.} \label{Spectra}
\end{figure}

Using the spectrometer (Shamrock 500i ruled grating with 1200 lines/mm from Andor) we also recorded the photoluminescence spectrum from 560 nm to 760 nm of both trapped diamonds and diamonds deposited on a glass coverslip. Some of the spectra obtained for different excitation power are shown in figure \ref{spectrum}. One can clearly identify the ZPL (Zero Phonon Line) of both NV$^-$ and NV$^0$ together with the broad PSB (Phonon Side Band) of the NV$^-$. A narrow peak due to Raman scattering is also visible at around 573 nm. Observations of a number of deposited diamonds showed that the relative intensities of the two ZPLs and of the Raman scattering peak vary from one particle to an other. The NV$^-$ to NV$^0$ population ratio is also seen to depend on the 532 nm excitation laser power for both trapped and deposited diamonds  as can be seen on the spectra of figure \ref{spectrum}. Note that the curves have been offset for clarity.

%Figure \ref{Spectra} shows the spectra obtained with NV centers in 10 $\mu m$ sized diamonds that are deposited on a glass coverslip (trace (a)) and trapped in the Paul trap (trace (b)) for two different excitation powers. 
%The count rates have been corrected for
%background light and normalized by the acquisition time and the excitation laser power.
These data suggest that there is no significant effect of the trap on the NV spectrum. However, since these measurements were done using two different particles, a more precise measurement still needs to be done in order to clearly show no effect of the trap on the NV spectrum. This could be done either by using diamonds samples where all particles display similar spectral properties, or by characterizing the same particle trapped and then deposited with a similar fiber deposition method as in \cite{Kuhlicke}.

\begin{figure}[ht!]
\centerline{\scalebox{0.35}{\includegraphics{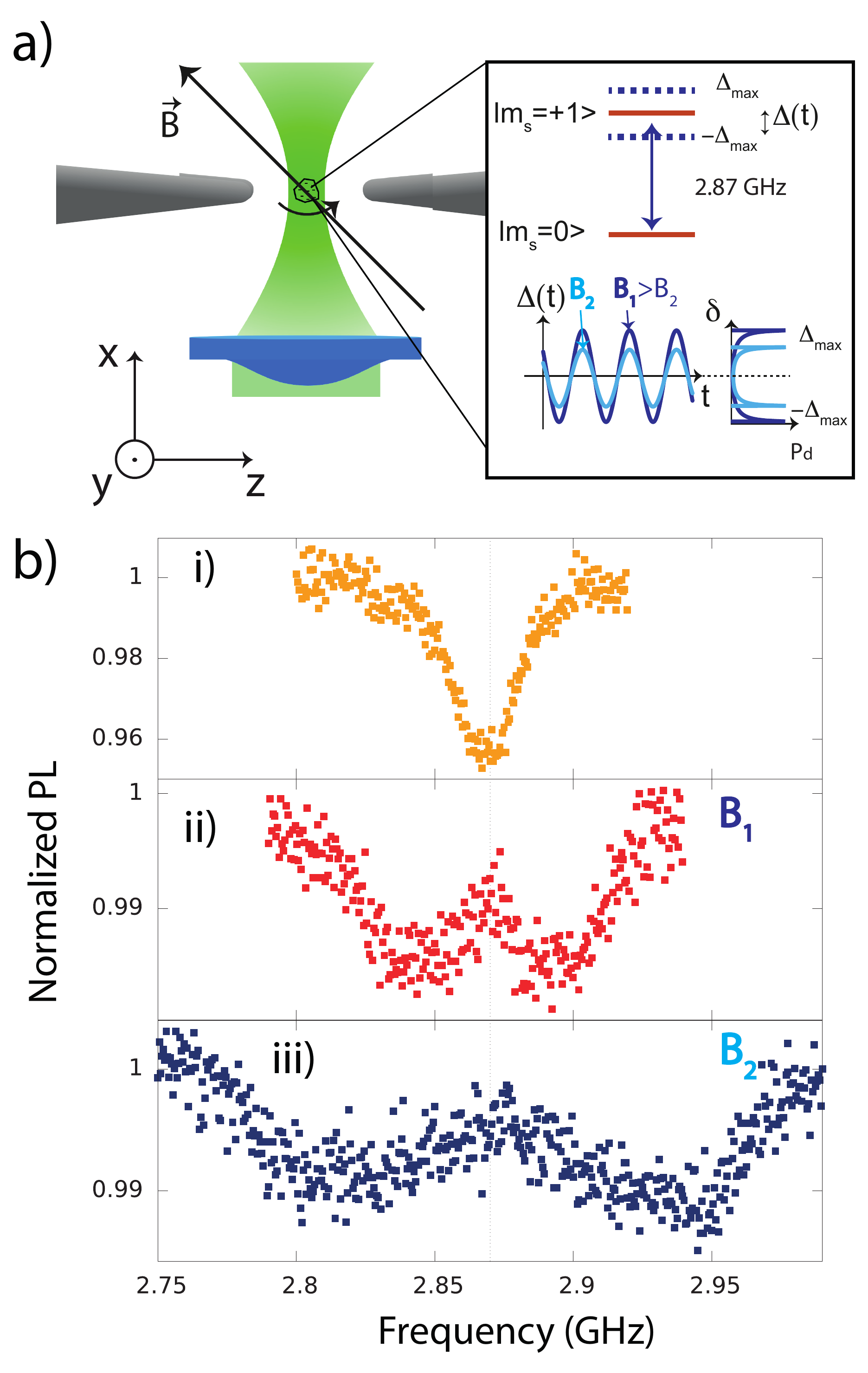}}}
\caption{a) Schematics showing the influence of the laser on the particle motion in the Paul trap and on the ESR spectrum. Under a magnetic field, a rotating single NV center's spin experiences a time varying frequency shift $\Delta(t)$ leading to a broader ESR following the sinusoidal distribution labelled $P_d$. b) Electron spin resonance (ESR) from levitating and rotating diamonds in a Paul trap. Trace i) was taken without externally applied magnetic field and traces ii) and iii) with magnetic fields in the range of 10 G and 30 G respectively.
}\label{ESR}
\end{figure}

\subsection{Electronic spin read-out from levitating diamonds} 

We can now probe the NV$^-$ centers' electronic spin transitions with the trapped diamonds.
The level structure of the NV$^-$ spin is also depicted in Figure~\ref{level}.
The NV$^-$ has two unpaired electrons so the ground state is a spin triplet. The degeneracy between the $m_s=0$ and $m_s=\pm 1$ manifolds, defined with respect to the NV center axis, is lifted by 2.87 GHz due to spin-spin interaction.
One of the important properties of the NV center is the photoluminescence rate dependency with respect to the electronic spin state. This is the result of a spin dependent non-radiative decay via the levels $^1A_1$ and $^1E$. As depicted in Figure~\ref{level}, when the spin is in the $m_s=\pm 1$ state, decay to these levels takes place. As a consequence, the PL rate drops. 
Under continuous microwave and optical excitation, scanning the frequency of a microwave tone around 2.87 GHz typically results in a change of the photoluminescence on the order of 10\% with a half-width at half maximum in the range of tens of MHz \cite{Gruber}.
Under a magnetic field, the degeneracy between the two $m_s=\pm 1$ states is lifted due to the Zeeman effect. For a single NV spin, two narrow lines will thus appear in the ESR spectrum, the frequency of which will depend upon the projection of the applied B-field onto the NV axis. This property means that NV centers can be used as sensitive room temperature magnetic field probes on microns lengths scales making NV centers very attractive for applications in magnetometry (see Ref.~\cite{rondin} for a review) and hybrid opto-mechanical schemes \cite{Rabl}. The NV centers can actually be found in the four equiprobable
$[\overline{1} \overline{1}1]$, $[1\overline{1}1]$, $[\overline{1},\overline{1},1]$, $[111]$ orientations within the diamond lattice.
In a monocrystal containing several NVs we thus expect to observe four Zeeman-splitted ESR lines for the two electronic transitions, because of the different magnetic field projections along the four NV axes.  

To observe the Electron Spin Resonance (ESR) using levitating diamond particles, a 28 $\mu m$ current carrying copper wire that lies 150 microns away from the trap center is used as an antenna, as depicted in Figure~1. 
We then noticed that the trapping parameters depend upon the distance between the antenna and the center of the trap. Simulations show that the stiffness of the trap increases when the distance decreases. As the antenna is brought close to the trap, the $q$ factor of the Mathieu equation thus increases and so the frequency and AC voltage are often adjusted to keep the $q$ factor well within the stability region. 
However, in the presence of the antenna, trapping is still critical and the particle is lost more often when sudden local changes in the pressure or nearby charges densities occur. 
This points towards a decrease of the potential depth due to the increasing asymmetry of the trap. This issue, also confirmed by numerical simulations, prevents us from acquiring data for more than half an hour and limits the signal to noise ratio of our ESR.

We perform two sets of experiments with two different particle sizes. In the first experiment, particles with mean sizes of 710 nm are used, whereas in the second, mean particles sizes of 9.6 $\mu$m will be used.  

Fig.~\ref{ESR}-b), trace i) shows the ESR spectrum obtained from diamonds with sizes of 710 nm.
As expected, we observe a decrease of the photoluminescence level with a minimum at the ESR transition frequency of 2.87 GHz without externally applied magnetic field.
No change in the ESR contrast was observed compared to when the diamonds are deposited on a quartz coverslip so the temperature of the diamond is close to room temperature.
%This compares favourably to optical traps, where the temperature of the levitating diamond is significantly modified by the trap, causing a reduction in the ESR contrast \cite{Hoang, Rahman, Neukirch}.  

For particle sizes below 1 $\mu$m the laser field sets the particle in motion on time scales that are much faster than the acquisition time of each ESR spectrum (seconds). In the presence of a magnetic field, a broadening of the ESR spectrum takes place as the magnetic field is increased since a single NV center's spin experiences a time varying magnetic field, as explained in the inset of Fig.~\ref{ESR} a). In reality, the ESR is actually the superposition of 8 sinusoidal distributions that can be displaced from the zero-field splitting ($\delta=0$). 
Qualitatively, a single NV center's spin experiences a time varying magnetic field $\Delta(t)$ leading to an ESR that follows a distribution $P_d$ that is sinusoidal, the amplitude of which (denoted $\Delta_{\rm max}$ in Fig.~\ref{ESR} a)) depends upon the magnetic field strength. In the inset, for simplicity, we describe an ESR for a transition $m_s=0$ to $m_s=+1$ and a single NV center whose rotation axis is perpendicular to the magnetic field. 
Besides, the efficiency of the micro-wave driving is time-modulated, leading to a further reshaping of the ESR profiles \cite{Horowitz}. ESR spectra are shown in Figure~\ref{ESR}-b), with traces ii) and iii) corresponding to increasing B-fields.
As anticipated, the experimental curves display a reduced contrast and broader linewidths demonstrating that the particle rotates in the magnetic field in accordance with the observed motion of the particle in transmission imaging. From the extremal frequency points of the graphs, we estimate the B-field to be 10 G for trace ii) and 30 G for trace iii). These values are derived from the projection maxima over the NV axis that goes through a maximal alignement with the magnetic field during its rotation.

\begin{figure}[ht!]
\centerline{\scalebox{0.12}{\includegraphics{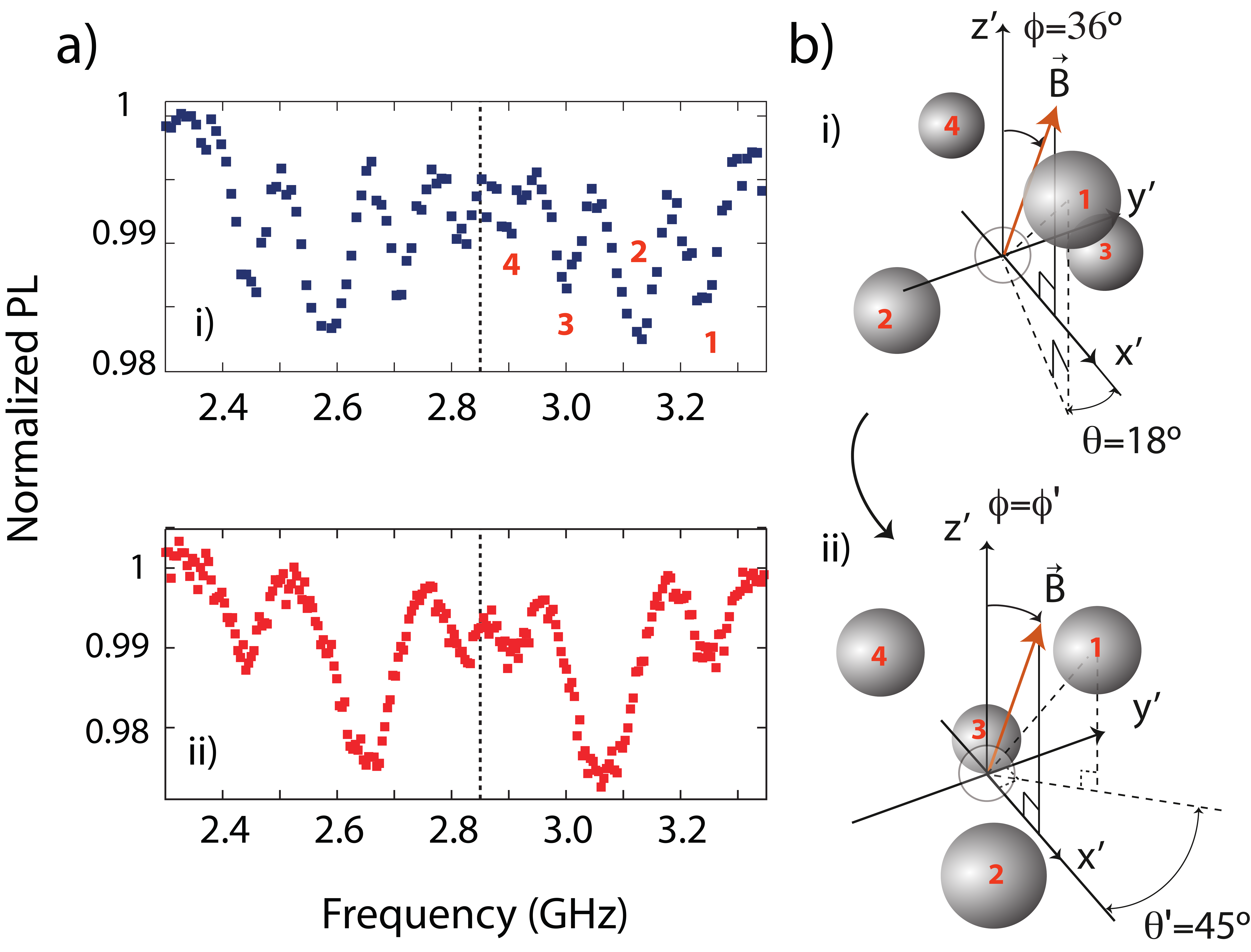}}}
\caption{a) Electron spin resonance (ESR) from a levitating diamond monocristal in a Paul trap. Traces i) and ii) correspond to two different confinement frequencies. b) NV centers orientations with respect to the applied magnetic field as deduced from the ESR spectra.}\label{ESR2}
\end{figure}

We now move to an experiment where microdiamonds with mean sizes 9.6 microns are injected in the trap. 
The corresponding ESR is shown Fig.\ref{ESR2}-a) trace i). It displays 8 resonances, corresponding to the projection of the two spin transitions $|m_s=0\rangle \rightarrow |m_s=\pm 1\rangle$ on the 4 orientations of the NV centers within the levitating diamond. Here, we adjusted the magnetic field angle and strength so that the spectral lines are equally separated. 
Compared to the results presented in Figure~\ref{ESR}, observing such a splitting demonstrates that the microdiamonds do not rotate during the course of the measurement and that we can stably trap single monocrystals for extended periods of time.
Let us note that the contrast of each line also varies due to differing alignements of the NV axes with respect to the microwave polarisation. 

Such angular stability was also shown using an optical tweezer with nanodiamonds in water \cite{Geiselmann} and under vacuum \cite{Hoang} by adjusting the polarisation of the trapping light. Here, the stability of the particle orientation can be understood by considering the relationship between its angular momentum and the torque applied by the electric field of the Paul trap.  Let us consider the rotation around the $y$ axis of an ellipsoidal particle in the trap. It can be shown that the angle $\alpha$ between the axis of the ellipsoid and the $z$ axis is given by the following equation :
%\begin{eqnarray}
%\ddot\phi- \sqrt{2} \omega_\phi \Omega \cos(\Omega t) \frac{\sin(2\phi)}{2}=0
%\end{eqnarray}

\begin{eqnarray}
\ddot\alpha- \sqrt{2} \omega_\alpha \Omega \cos(\Omega t) \frac{\sin(2\alpha)}{2}=0
\end{eqnarray}

where $\omega_z$ is the axial angular frequency of the harmonic pseudo-potential for the center of mass, $I_{yy}$ is the moment of inertia about the $y$ axis, 

$$\rm S_I=\frac{3}{S} \iint \left( z^2-x^2 \right) dS$$

 with $x$ and $y$ defined in the reference frame of the ellipsoid and $S$ is the surface of the ellipsoid. This relation is deduced from integrating the torque applied by the electric field on the surface charges and factorizing the $\alpha$ dependency. In the limit of small rotations around $\alpha=0$ or $\alpha=\frac{\pi}{2}$, this equation is similar to the one that governs the dynamics of the center of mass of the particle in the trap. It thus leads to a harmonic pseudo-potential for the angle $\alpha$, here with angular frequency $\omega_\alpha$. 
It thus leads to a harmonic pseudo-potential for the angle for the rotation about the $x$ and $y$ axis. Moreover since neither the levitating particle nor the trap setup are actually rotationally symmetric about the $z$ axis, we can also expect a confinement for the rotation about the $z$ axis. 

Using the ESR spectra, one can deduce the angle of the diamond monocristal orientation with respect to the magnetic field and the value of the magnetic field.  We write $(\theta,\phi)$ the azymutal and polar angles respectively for the coordinates of the NV orientation labelled $1$ in the sketch b)-i) in Figure~\ref{ESR2}. 

In the experiment where micronsize particles are injected in the trap, four equidistant ESR peaks are obtained for both positive and negative frequencies in Fig.~\ref{ESR2} a).
We consider only the positive part of the spectrum. Writing $\vec{x}_1=[1,1,1]$, $\vec{x}_2=[\overline{1},1,1]$, $\vec{x}_3=[1,\overline{1},1]$ and $\vec{x}_4=[\overline{1},\overline{1},1]$  for the directions of the NV axes and $\vec{B}=B(\cos\theta \sin\phi, \sin\theta \sin\phi, \cos\phi)$, with 
$\phi=]0,\pi/4[$, the equidistance between the peaks lead to the condition $\tan \theta=2 +n \pi$.
This also gives the following equations for the projections of the four NV on the B field 
\begin{eqnarray}
\nonumber
\vec{x}_1\cdot \vec{B}&=&(\frac{3}{2}\sin\theta \sin\phi +\cos\phi)B \\
\nonumber
\vec{x}_2\cdot \vec{B}&=&(\frac{1}{2}\sin\theta \sin\phi +\cos\phi)B\\
\nonumber
\vec{x}_3\cdot \vec{B}&=&(-\frac{1}{2}\sin\theta \sin\phi +\cos\phi)B\\
\nonumber
\vec{x}_4\cdot \vec{B}&=&(-\frac{3}{2}\sin\theta \sin\phi +\cos\phi)B.
\end{eqnarray}
%where $\gamma_e=2.8$~MHz/G is the gyromagnetic factor of the electron. 
The condition $\vec{x}_i\cdot \vec{B}>0$ imposes $n=0$, which implies that $\theta=\arctan 2$.

We can deduce the value of the magnetic field and the angle $\phi$ by measuring the Zeeman frequency shifts of the levels corresponding to orientations 1 and 2 with respect to the zero-field splitting line. 
These are $\omega_1/2\pi=0.37$ GHz and $\omega_2/2\pi=0.25$ GHz. Writing $r=\omega_1/\omega_2$, we obtain 
the condition
\begin{eqnarray}
\tan\phi=\frac{r-1}{\frac{1}{2}\sin\theta(3-r)},
\end{eqnarray}
giving $\phi=36$ degrees. We then obtain the magnitude of the B field to be 80 G knowing the gyromagnetic factor of the electron $\gamma_e=2.8$~MHz/G. The resulting NV centers' structure with respect to the B-field axis is depicted in Fig.~\ref{ESR2}-b-i). In this figure, $\theta=\arctan(2)-\pi/4$, i.e. about $18 ^{\circ}$ with respect to the orientation $1$ because the directions of the axes 2 and 3 are inverted for more readibility and the
tetrahedron is centered at the origin in a coordinate frame $(x',y',z')$ for which the projection of the B-field is 0 along $x'$.
%
%In Fig.\ref{ESR2}-b), the particle has been rotated after changing the frequency of the trap. 
%The extremal ESR peak positions do not shift, meaning that, in the NV frame, $\vec{x}_4\cdot (\gamma_e \vec{B})=\vec{x}_4\cdot (\gamma_e \vec{B}')$, where $\gamma_e=2.8$~MHz/G is the gyromagnetic factor of the electron and $B'$ is the B field seen by the rotated particle.
%The two central peaks merge, which implies that $\vec{x}_3\cdot (\gamma_e \vec{B}')=\vec{x}_2\cdot (\gamma_e \vec{B}')$.
%The first condition implies that $\phi'=\phi=36$ degrees while the second implies that $\theta'=0$.
%Let us note that there are degeneracies since $n\pi/2$ rotations with the polar angle $\theta$ ($n$ an integer number) does not change the ESR spectrum while the diamond may have rotated. 

%One finds that 
%$\theta=\arctan(2)-\pi/4$, i.e. about $18 ^{\circ}$ with respect to the orientation $1$. We then deduce the angle $\phi$ to be 36 degrees and the magnetic field amplitude to be 80 G. The resulting NV centers' structure with respect to the B-field axis is depicted in Fig.~\ref{ESR2}-b-i)
%\footnote{The directions of the axes 2 and 3 are inverted for more readibility and the
%tetrahedron is centered at the origin in a coordinate frame $(x',y',z')$ for which the projection of the B-field is 0 along $x'$.}.

When tuning the trapping frequency from $\Omega/2\pi=3.3$ kHz to $\Omega/2\pi=2.6$ kHz
we observe a change in the phase contrast image which points to a deterministic change in the particle angle.
We attribute the rotation to the presence of residual electric fields on the needles. 
Here, the patch potentials on the needle generate an angle/position dependent residual field in the vicinity of the particle, which when integrated over the charges of the particle will result in an angle-dependent torque.
Residual electric fields due to patch potentials typically shift the center of mass of ions when the trap voltage or frequency is changed, because of the change in the relative weight between the Paul trap potential and this extra electric potential.
If instead such angle-dependent torque is applied to the particle, this will shift the stable angular position.
We observe a change in the phase contrast image which indeed points towards a deterministic rotation of the particle on itself.
Trace ii) displays an ESR spectrum taken with the same particle as for trace i), but with this increased confinement.  %A displacement of the frequency of the two central lines corresponding to orientations 2 and 3 of the $|m_s=0\rangle$ to $|m_s= 1\rangle$ transitions is observed while lines 1 and 4 do not shift significantly. 

%We can thus deduce a relationship between the orientations $(\theta',\phi')$ in this new stable position of the diamond, using the conservation of the projection for orientations 1 and 4 and find that $\phi'=\phi$. The merging of the two central peaks yields $\theta'=45^{\circ}$. Let us note that there are degeneracies since $n\pi/2$ rotations with the polar angle $\theta$ ($n$ an integer number) does not change the ESR spectrum while the diamond may have rotated..
The extremal ESR peak positions do not shift, meaning that, in the NV frame, $\vec{x}_4\cdot (\gamma_e \vec{B})=\vec{x}_4\cdot (\gamma_e \vec{B}')$, where %$\gamma_e=2.8$~MHz/G is the gyromagnetic factor of the electron and 
$B'$ is the B field seen by the rotated particle.
The two central peaks merge, which implies that $\vec{x}_3\cdot (\gamma_e \vec{B}')=\vec{x}_2\cdot (\gamma_e \vec{B}')$.
The first condition implies that $\phi'=\phi=36$ degrees while the second implies that $\theta'=0$.
Let us note that there are degeneracies since $n\pi/2$ rotations with the polar angle $\theta$ ($n$ an integer number) does not change the ESR spectrum while the diamond may have rotated. 

We conclude that the diamonds necessarily rotated around the vertical axis $z'$ after this change in the trap parameters, as depicted in Fig.~\ref{ESR2}-b ii). In this figure, $\theta'=45 ^{\circ}$ for the same reason of readibility than on Fig.~\ref{ESR2}-b i).

\subsection{Conclusion}
We observed electron spin resonances from NV centers with single diamond monocrystals levitating in a Paul trap. 
This experiment is realized in a unique regime where the trap does not impact the photo-physical properties of NV centers. Our results furthermore show angular stability of the charged diamonds over time scales of minutes, 
a necessary step towards spin-controlled levitating particles. 
Ramsey spectroscopy or electric field noise measurements \cite{Jamonneau} will then be implemented to assess the applicability of nanodiamonds containing NV$^-$ centers in ion traps for quantum sensing. 
This system can for instance be a potentially useful tool for vectorial magnetometry, and enable the observation of quantum geometric phases \cite{Maclaurin} and be used as high precision multi-axis rotational sensor \cite{Maclaurin, Ledbetter}. 
Using a more confining trap combined with electrospray ionisation for loading, will enable selection of particles with higher charge to mass ratios and thus pave a path towards efficient single spin opto-mechanical schemes \cite{Scala, Rabl, arcizet, DelordAngular}. 

\begin{acknowledgements}
We would like to acknowledge fruitful discussions with the optics team at LPA and with V. Jacques, L. Rondin, F. Treussard, J. -F. Roch, Q. Glorieux, F. Dubin, L. Guidoni, J.-P Likforman, and B. Fox. We also thank A. Denis and the team of D. Courtiade for their help in the initial phase of the experiment. 
This research has been partially funded by the French National Research Agency (ANR) through the project SMEQUI.
\end{acknowledgements}

\end{document}